\documentclass[letterpaper,12pt]{article}
\usepackage{fullpage,natbib}

\usepackage{graphicx}
\usepackage{amsmath,amssymb}
\usepackage{bm}
\usepackage{multirow}
\usepackage{color}
\usepackage[usenames,dvipsnames,svgnames,table]{xcolor}

\title{Core-scale solute transport model selection using Monte Carlo analysis}
\author{Bwalya Malama, Kristopher L. Kuhlman, and Scott C. James}

\begin{document}
%%\begin{linenumbers}
\maketitle

%%\authors{}
%%\affil{Sandia National Laboratories, Carlsbad, New Mexico}
%%
%%\author{Scott C. James}
%%\affil{Exponent, Inc., Irvine, California}

\begin{abstract}
Model applicability to core-scale solute transport is evaluated using breakthrough data from column experiments conducted with conservative tracers tritium $\left(^3\mathrm{H} \right)$ and sodium-22 $\left(^{22} \mathrm{Na} \right)$, and the retarding solute uranium-232 $\left(^{232}\mathrm{U}\right)$. The three models considered are single-porosity, double-porosity with single-rate mobile-immobile mass-exchange, and the multirate model, which is a deterministic model that admits the statistics of a random mobile-immobile mass-exchange rate coefficient. The experiments were conducted on intact Culebra Dolomite core samples. Previously, data were analyzed using single- and double-porosity models although the Culebra Dolomite is known to possess multiple types and scales of porosity, and to exhibit multirate mobile-immobile-domain mass transfer characteristics at field scale. The data are reanalyzed here and null-space Monte Carlo analysis is used to facilitate objective model selection. Prediction (or residual) bias is adopted as a measure of the model structural error. The analysis clearly shows single- and double-porosity models are structurally deficient, yielding late-time residual bias that grows with time. On the other hand, the multirate model yields unbiased predictions consistent with the late-time $-5/2$ slope diagnostic of multirate mass transfer. The analysis indicates the multirate model is better suited to describing core-scale solute breakthrough in the Culebra Dolomite than the other two models.
\end{abstract}

%\begin{article}

\section{Introduction}
During the last 30 years, significant effort has been expended to understand contaminant transport in fractured rock \citep{huyakorn1983, sun2003} due in part to the necessity to evaluate site suitability for nuclear waste disposal. Contaminant transport in fractured rock is of common concern to regulators and stakeholders at nuclear waste disposal sites because off-site contaminant migration could impact groundwater resources. Modeling contaminant transport in fractured rock is challenging due to the complex and inherently heterogeneous nature of the transport domain, and the multitude of physical and chemical processes controlling contaminant interaction with the host rock. This has led to a development of several potentially competing conceptualizations of the transport environment \citep{genuchten1989,zheng2010}. Model selection is typically based on subjective expert judgment. Hence, there is a need for objective criteria for selecting physically-based models that best describe observed transport behavior and provide minimal predictive uncertainty.

In this work, we present a criterion for selecting between competing models for describing transport at the core scale. Three models are considered: the single-porosity model; the traditional double-porosity model with single-rate mobile-immobile domain mass exchange \citep{genuchten1989, gamerdinger1990}, and; a double-porosity model with multiple rates of mobile-immobile-domain mass exchange controlled by a random mass transfer coefficient \citep{haggerty1995,haggerty1998}. We refer to the traditional double-porosity model as simply the double-porosity model, and to the model with multiple rates of mass exchange as the multirate model following \citet{haggerty1995,haggerty2000} and \citet{meigs2000}. In the multirate model, the mass transfer coefficient is a random variable, not a single deterministic parameter. This conceptualization reflects spatial, not temporal, variability (due to heterogeneity, i.e., multiple types and scales of porosity). The probability density function of the transfer coefficient gives the probability that a mobile-immobile interface (assumed to be randomly distributed in space), encountered by a particle along its trajectory through the transport domain, has a particular mass transfer coefficient value.

The three models are used to analyze breakthrough data collected in core-scale laboratory experiments \citep{lucero1998} using conservative tracers tritium $\left(^3 \mathrm{H}\right)$ and sodium-22 $\left(^{22} \mathrm{Na}\right)$, and the retarding tracer uranium-232 $\left(^{232} \mathrm{U} \right)$. The experiments analyzed herein were performed on a rock core collected from a formation known to exhibit multiple types and scales of rock-matrix porosity. Previous analysis of the experimental data with single- and double-porosity models by \citet{lucero1998} yielded poor model fits to these data due to the inability of the two models to describe the long-tailing behavior of conservative solutes. The multirate model has been shown to properly describe this behavior in breakthrough data obtained in field-scale tracer tests \citep{meigs01, haggerty2001, mckenna2001}. It is applied herein for the first time to core-scale breakthrough data to demonstrate multirate mass-transfer effects are observable at this scale.

Null-space Monte Carlo analysis (NSMC) is used to evaluate model prediction uncertainty for each of the three model based on breakthrough data. It yields multiple sets of parameters that calibrate the model \citep{tonkin2007,tonkin2009,james2009,gallagher2007}, leading to multiple realizations of model fits to data at parameter estimation optimality. By prediction uncertainty we mean the variance and bias of the ensemble of these model-prediction realizations relative to observed behavior. Variance describes the scatter of realizations about mean behavior, while the residuals bias associated with each data point at optimality over all NSMC realizations provides a measure of the systematic departure of predicted from observed behavior. This work presents the first use of residual bias in the solute transport literature as a criterion for model selection.

\section{The multirate transport model}
The multirate model is based on the traditional double porosity model where the transport domain is conceptualized as comprising two overlapping continua, namely the mobile (advective or fracture porosity) and immobile (diffusion-dominated matrix porosity) domains. Unlike the traditional double porosity model where a single deterministic constant is used to characterize mobile-immobile-domain mass exchange, a random variable is used in the multirate model. Using this conceptual approach, the governing equation for transport of a sorbing radionuclide in the mobile domain \citep{haggerty1995,haggerty1998} is given in nondimensional form
\begin{equation}
\label{eqn:gov-dimless}
\frac{\partial C}{\partial T} + \int_0^\infty \beta(\omega_D)\left(\frac{\partial C_\mathrm{im}}{\partial T} + \lambda_D C_\mathrm{im} \right) \mathrm{d}\omega_D = \frac{1}{\mathrm{Pe}} \frac{\partial^2 C}{\partial X^2} - \frac{\partial C}{\partial X} - \lambda_D C,
\end{equation}
where $C = c/C_c$, $C_\mathrm{im} = c_\mathrm{im}/C_c$, $X = x/L_c$, $T = t/T_c$, $c$ and $c_\mathrm{im}$ are mobile- and immobile-phase solute concentrations [M~L$^{-3}$], $x$ and $t$ are space-time coordinates, $C_c$, $L_c$, and $T_c$ are characteristic concentration, length, and time,  $\lambda_D = \lambda T_c$, $\lambda$ is the first-order radioactive decay constant [T$^{-1}$], $\omega_D = \omega T_c$ is the dimensionless first-order mass-transfer rate coefficient (Damk\"{o}hler-I number), $\beta(\omega_D)=\beta_T p(\omega_D)$ is the rock matrix point capacity ratio, $\beta_T=\phi_\mathrm{im} R_\mathrm{im} / \phi_\mathrm{m} R_\mathrm{m}$ is the dimensionless rock-matrix total capacity ratio, $p(\omega_D)$ is the probability density function (pdf) of $\omega_D$, $\mathrm{Pe} = L_c/\alpha_L$ is the P\'{e}clet number, $\alpha_L$ [L] is the longitudinal dispersivity, $\phi_\mathrm{m}$ and $\phi_{\mathrm{im}}$ are the mobile- and immobile-domain porosities, and $R_\mathrm{m}$ and $R_\mathrm{im}$ are the mobile- and immobile-domain retardation factors.

The dimensionless governing equation for immobile domain transport is
\begin{equation}
\label{eqn:mass-exD}
\frac{\partial C_\mathrm{im}}{\partial T} = \omega_D(C - C_\mathrm{im}) - \lambda_D C_\mathrm{im},
\end{equation}
the lumped-parameter formulation of immobile-domain mass transport.

The transport equations are solved subject to the initial condition
\begin{equation}
\label{eqn:initialD}
C(X,T=0) = C_\mathrm{im}(T=0) = C_0,
\end{equation}
indicating initial equilibrium between mobile and immobile-domain concentrations. The boundary condition at $X=0$ is 
\begin{equation}
\label{eqn:upstream-bcD}
\left.\left(\frac{A}{L_c}\frac{\partial C}{\partial X} + B C \right)\right|_{X = 0} = C_\mathrm{inj} (T),
\end{equation}
where $C_\mathrm{inj}$ is a normalized injection concentration and $A$ [L] and $B$ are parameters to specify the $X=0$ boundary condition type ($A = 0$ and $B=1$ correspond to a Dirichlet boundary condition, while $A=-D/v$ and $B=1$ correspond to a Robin boundary condition). The downstream boundary condition is
\begin{equation}
\label{eqn:downstream-bcD}
\lim_{X\rightarrow \infty} \left(-\frac{1}{\mathrm{Pe}}\frac{\partial C}{\partial X} + C\right) = 0,
\end{equation}
indicating zero solute flux infinitely far downstream.

The solution to (\ref{eqn:gov-dimless})--(\ref{eqn:downstream-bcD}) is obtained on a semi-infinite domain $0\leq X < \infty$ as a simplification and limiting case of the finite domain considered by \citet{haggerty1995,haggerty1998}. It is given by
\begin{equation}
\label{eqn:c-sol}
\bar{C} (X) = \left(\frac{\bar{C}_\mathrm{inj} - B \bar{C}_p}{B + uA/L_c}\right) e^{uX} + \bar{C}_p,
\end{equation}
where the overbar indicates the Laplace transform, $s$ is the Laplace transform parameter, $u = \left(1-\sqrt{1 + 4\bar{f}_1/P}\right)P/2$, $\bar{C}_p = C_0/(s + \lambda_D)$, $\bar{f}_1(\lambda_D) = (s+\lambda_D) \bar{f}_0(\lambda_D)$, $\bar{f}_0(\lambda_D) = 1 + \beta_T \bar{g}(\lambda_D)$, and
\begin{equation}
\label{eqn:memorykernel}
  \bar{g}(\lambda_D) = \int_0^\infty \frac{\omega_D p(\omega_D)}{s + \lambda_D + \omega_D} \;\mathrm{d}\omega_D.
\end{equation}
The function $\bar{g}(\lambda_D)$ is the Laplace transformed memory function of \citet{haggerty2000}. For single porosity $\bar{g}(\lambda_D) \equiv 0$, whereas for double porosity with single-rate mass transfer $\bar{g}(\lambda_D) = \omega_D/(s + \lambda_D + \omega_D)$. The inverse Laplace transform of (\ref{eqn:c-sol}) is obtained using the \citet{dehoog1982} algorithm. For all results reported herein, $C_c = c_\mathrm{inj}$ is the injection concentration, $L_c$ is core length, and $T_c = L_c/v_R$, where $v_R = v/R_\mathrm{m}$ and $v$ is the average linear velocity [L~T$^{-1}$].

\subsection{Mass-Transfer Coefficient Distribution}
To evaluate the memory kernel $\bar{g}(\lambda_D)$ numerically, the probability density function $p(\omega_D)$ must be specified. All valid probability density functions are admissible in the computation of the memory function, including single-parameter distributions such as the power-law used by \citet{haggerty2000} and \citet{schumer2003}. However, single-parameter distributions may not lead to improved multirate model predictions of breakthrough behavior compared to the single-rate mass transfer model. Here, without loss of generality, we use the lognormal distribution because several key geological properties appear to approximately follow this distribution \citep{haggerty1998}, including hydraulic conductivity \citep{neuman1982, hoeksema1985} and grain size \citep{buchan1993}. Other equally valid examples of distributions that have been used in the literature to characterize the mobile-immobile mass transfer coefficients are summarized in \citet{haggerty2000}. Using any of these models with two or more parameters, would likely yield multirate models that outperform the single-porosity and single-rate double-porosity models.

The standard two-parameter lognormal distribution for $\omega_D \in [0,\infty)$ was used by \citet{haggerty1995,haggerty1998}. For the case where physical bounds exist $\omega_D \in [\omega_{D,\mathrm{min}},\omega_{D,\mathrm{max}}]$, it may be more appropriate to use the random variable  $\hat{\omega}_D = \left(1/\omega_D - 1/\omega_{D,\mathrm{max}}\right)^{-1} - \omega_{D,\mathrm{min}}$, where $\omega_{D,\mathrm{min}}$ and $\omega_{D,\mathrm{max}}$ are the minimum and maximum physically allowable $\omega_D$ values. The pdf of $\hat{\omega}_D$ is
\begin{equation}
p(\hat{\omega}_D) = \frac{1}{\hat{\omega}_D \hat{\sigma} \sqrt{2\pi}} \exp\left[-\left( \frac{\log(\hat{\omega}_D)-\hat{\mu}}{\hat{\sigma}\sqrt{2}}\right)^2\right],
\end{equation}
where $\hat{\mu}$ and $\hat{\sigma}$ are the mean and standard deviation of $\log(\hat{\omega}_D)$. This is the lower- and upper-tail-truncated lognormal distribution, which is a more plausible distribution when there are physical limits on permissible Damk\"{o}hler-I numbers. These physical limits may be estimated from data. In the limit as $\omega_{D,\mathrm{min}}\rightarrow 0$ and $\omega_{D,\mathrm{max}}\rightarrow \infty$, $\hat{\omega}_D \rightarrow \omega_D$, and $p(\hat{\omega}_D)$ degenerates to the standard two-parameter lognormal distribution. We set $\omega_D \in [0,1000]$, loosely based on the work of \citet{haggerty1995}, where $\omega_D = 100$ was suggested as the limit for significant multirate mass transfer.

\section{Application to Core-scale Breakthrough Data}
\label{sec:pest}
\begin{figure}[!hbtp]
  \centering
  \includegraphics[width=0.70\textwidth]{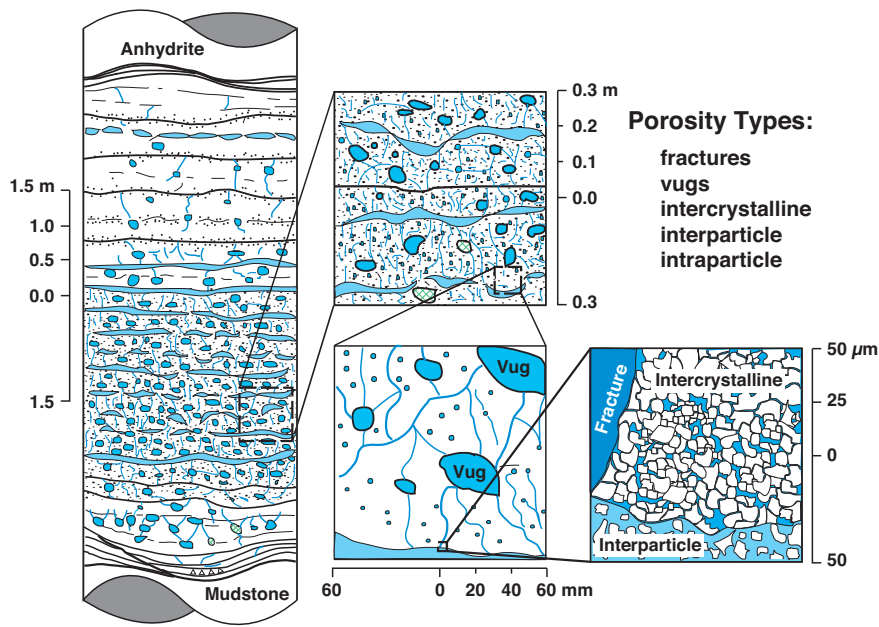}
  \caption{Different types and scales of Culebra Dolomite porosity.
    Anhydrite and mudstone of adjacent Rustler members act as
    confining layers.}
  \label{fig:porosities-diagram}
\end{figure}

\begin{figure}[!hbtp]
  \centering
  \includegraphics[width=0.80\textwidth]{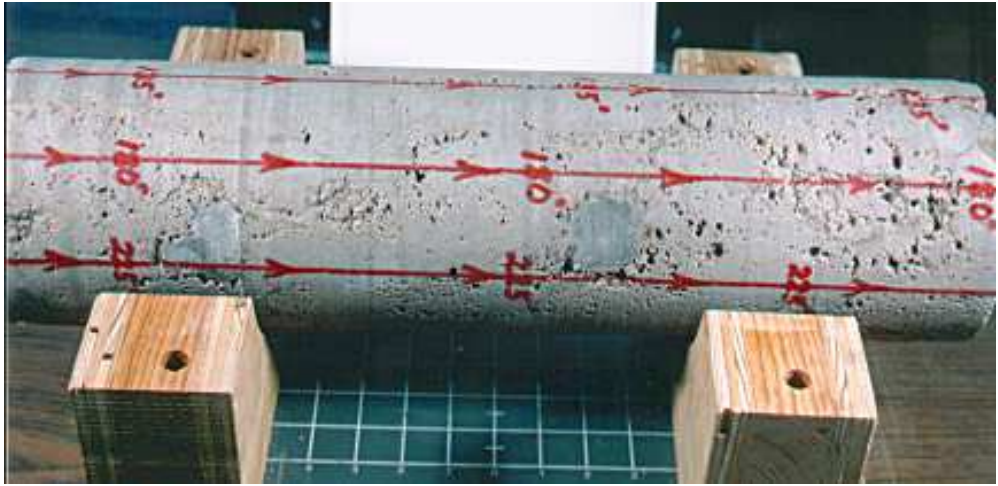}
  \caption{Culebra Dolomite horizontal core B showing vuggy porosity,
    fractures, and vug-filling calcite.  Foreground grid marks are
    inches.}
  \label{fig:corephotos}
\end{figure}

The data considered here were collected in a series of column experiments conducted on five intact cores (denoted A through E) of the Culebra Dolomite as reported by \citet{lucero1998}. The Culebra Dolomite member of the Rustler formation of the Permian Basin in southeastern New Mexico is known to exhibit several categories and scales of porosity \citep{holt1997} including inter-crystalline, inter-particle, fracture, and vuggy porosities (Figure~\ref{fig:porosities-diagram}). The multiple types and scales of porosity are also clearly observable in Culebra Dolomite cores (Figure~\ref{fig:corephotos}). The only breakthrough data analyzed in this work were collected on the B core for the conservative tracers $^3$H and $^{22}$Na, and the retarding tracer $^{232}$U. Core B, pictured in Figure~\ref{fig:corephotos}, was selected because its length-to-diameter ratio (50.9 cm to 14.5 cm) ratio was such that boundary effects can be neglected, thus permitting the use of the analytical solution developed for a 1D semi-infinite $(0\leq x <\infty)$ transport domain. Dry bulk density $\rho_\mathrm{bulk} = 2400$ kg/m$^3$ and total porosity $\phi_\mathrm{T} = 0.14$ were determined by standard laboratory methods \citep{lucero1998}. Additional details on experiment setup, solute injection, flow rates, and effluent analysis, are available in \citet{lucero1998} and are not repeated here.

\begin{figure}[!hbtp]
  \includegraphics[width=\textwidth]{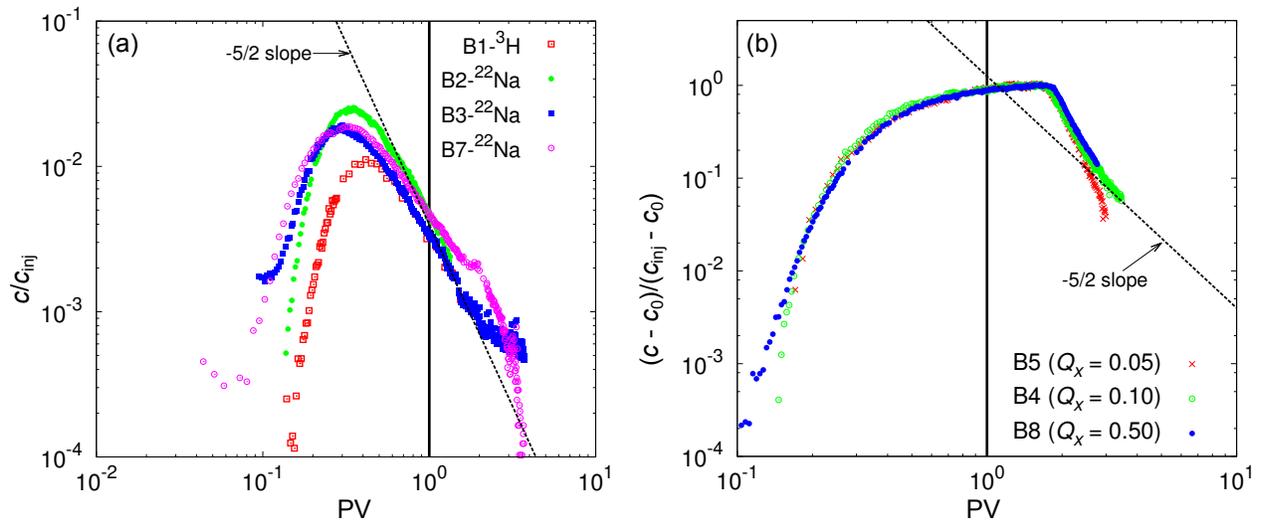} 
  \caption{Normalized concentrations plotted against PV for (a)
    short-injection-pulse tests B1, B2, B3, and B7, and (b) long
    constant-concentration-injection tests B4, B5, and B8. Vertical
    line marks one PV calculated using total porosity. $Q_x$ in (b) is
    volume flow rate in ml/min.}
  \label{fig:coreB}
\end{figure}

\begin{figure}[!hbtp]
  \includegraphics[width=\textwidth]{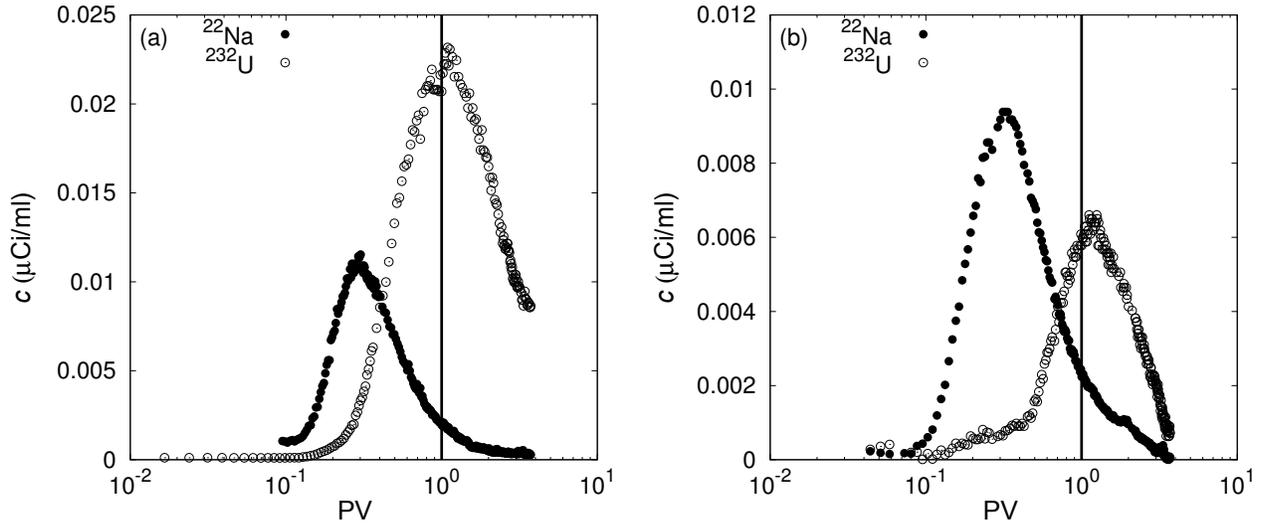} 
  \caption{Concentrations plotted against PV for $^{22}$Na
    (conservative) and $^{232}$U (retarding) in tests B3 (a) and B7
    (b). Vertical line marks one PV calculated using total porosity.}
  \label{fig:BNaU}
\end{figure}

Figure \ref{fig:coreB} shows normalized concentrations plotted against pore volume (PV) computed using $\phi_\mathrm{T}$. Solute injection pulses were longer in duration for tests shown in Figure~\ref{fig:coreB}(b) than for those in Figure~\ref{fig:coreB}(a). Plotting data on a log-log scale as in Figure~\ref{fig:coreB}(b) clearly shows that the effluent was not collected for a sufficiently long time to completely reveal the late-time tracer behavior. A long breakthrough tail is characteristic of mobile-immobile-domain mass transfer for conservative tracers. Despite this shortcoming, the data can be used to assess the performance of the three models. The data in Figure~\ref{fig:coreB}(a) show early breakthrough for both conservative tracers \citep{lucero1998}, suggesting the occurrence of preferential flow in an advective porosity that is significantly smaller than the total core porosity $\phi_\mathrm{T}$. Breakthrough data for $^{232}$U are shown in Figure~\ref{fig:BNaU} ($^{22}$Na data from the same test are included for comparison). $^{232}$U breakthrough clearly occurs much later than $^{22}$Na because the former sorbs onto the Culebra Dolomite. Peak $^{232}$U concentration arrival occurs around 1 PV, about four times later than  $^{22}$Na. Using the single-porosity model, \citet{lucero1998} estimated the $^{232}$U retardation factor to be 4.5 and 3.7, from B3 and B7 data, respectively. For the dual-porosity model, they obtained mobile- and immobile-zone retardation factor values of $\left\{R_\mathrm{m} =1.14, R_\mathrm{im}=65.4\right\}$ and $\left\{R_\mathrm{m} =4.35, R_\mathrm{im} = 1.00\right\}$, from B3 and B7 data, respectively. The value of $R_\mathrm{im}=65.4$ appears to be an error in recording the estimated value.

\subsection{Parameter Estimation}
To estimate model parameters we let $\bm{c}_\mathrm{obs}$ be the breakthrough data vector, $\bm{c}_\mathrm{cal}(\bm{\theta})$ the model-calculated concentrations vector, and $\bm{\theta}$ the vector of estimated model parameters. For $^3$H and $^{22}$Na, $\bm{\theta} = \left(\phi_\mathrm{m}, \alpha_L, \mu, \sigma, t_\mathrm{inj} \right)$, whereas for $^{232}$U, $\bm{\theta} = \left(R_\mathrm{m}, R_\mathrm{im}, \mu, \sigma \right)$. Injection pulse concentration $(c_\mathrm{inj})$ was fixed for tests B1, B2, B3, and B7, but was estimated for tests B4, B5, and B8. Increased test durations for B4, B5, and B8 made it more difficult to maintain constant injection concentrations over prolonged test periods, resulting in injection concentrations that varied appreciably with time \citep{lucero1998}. Since this temporal variability is not incorporated explicitly into the solution, and its functional form in unknown, the injection concentrations for tests B4, B5, and B8 are treated as unknown constants and are estimated from breakthrough data. Initial concentration $(c_0)$ was fixed for all tests and was determined from effluent concentration values measured prior to solute injection. The truncated lognormal distribution $(\omega_D \in [0,1000])$ was used to describe the mass-transfer coefficient distribution. The advective porosity $(\phi_\mathrm{m})$, dispersivity, and the injected pulse ($t_\mathrm{inj}$) duration were estimated with the multirate model for $^{22}$Na data and used as fixed input parameters when estimating the retardation factor and $\omega_D$ distribution parameters from $^{232}$U data. Distribution parameters were also estimated for $^{232}$U because $\omega_D$ is a function of the tracer-specific molecular diffusion coefficient.

\begin{figure}[!hbtp]
  \includegraphics[width=\textwidth]{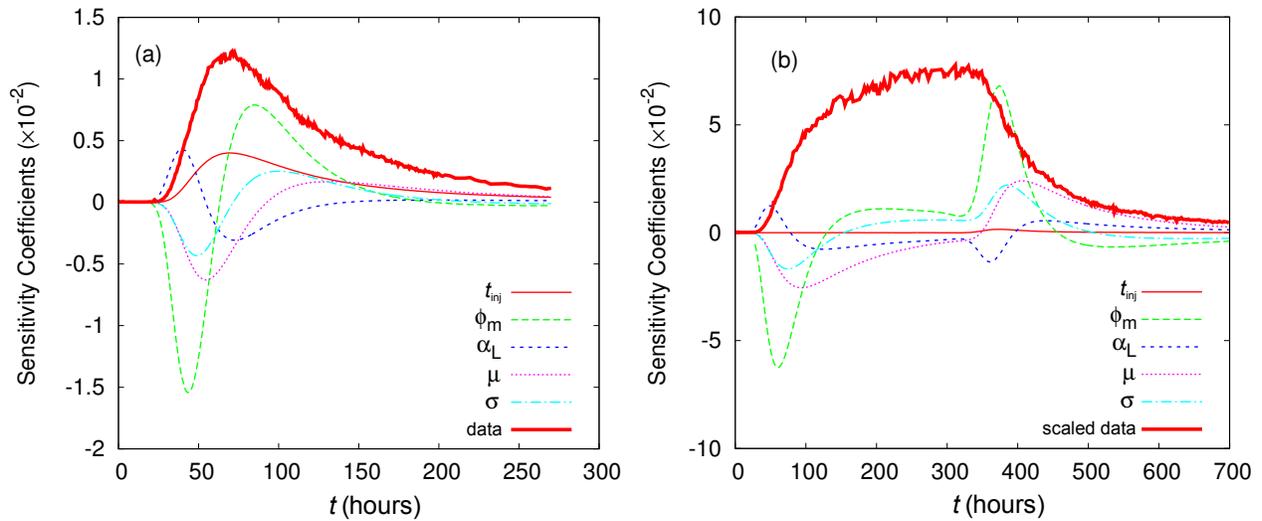} 
  \caption{Breakthrough concentration sensitivities to estimated
    multirate model parameters for (a) short- (B2) and (b) long-pulse
    (B4) $^{22}$Na tests. Concentration data are included for
    reference (those in (b) are scaled by 0.5).}
  \label{fig:sen1}
\end{figure}

\begin{figure}[!hbtp]
  \centering
  \includegraphics[width=0.49\textwidth]{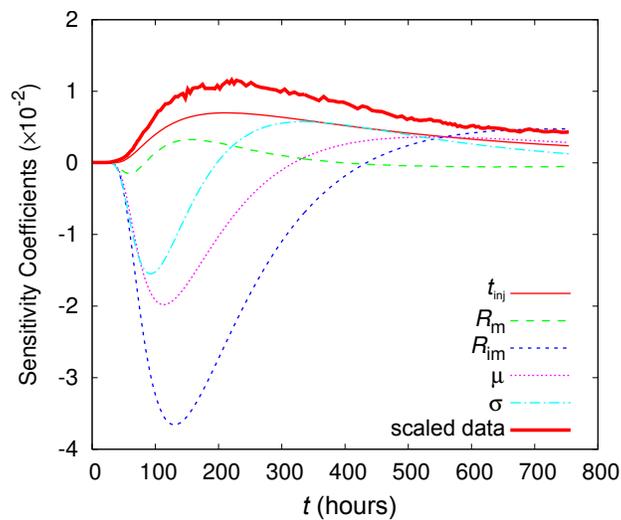}
  \caption{Breakthrough concentration sensitivities to estimated
    multirate model parameters for $^{232}$U in test B3.}
\label{fig:sen2}
\end{figure}

We examine model sensitivity coefficients to determine whether all model parameters are estimable from available data. Sensitivity coefficients are derivatives of model-predicted effluent concentrations with respect to model parameters, which are elements of the Jacobian matrix ($\mathbf{J}$).  They provide a measure of parameter identifiability, because the determinant of $\mathbf{J}^T\mathbf{J}$ must be sufficiently larger than zero to be estimable from data \citep{ozisik2000}. Small sensitivity coefficients imply $|\mathbf{J}^T\mathbf{J}| \approx 0$ and the inverse problem is ill conditioned. Here, sensitivity coefficients were estimated with PEST using central differences, and their variation with time is shown in Figure~\ref{fig:sen1} for (a) short (B2) and (b) long (B4) solute injection pulses. The sensitivities are sufficiently larger than zero to permit estimation of all parameters from breakthrough data. The coefficients are also linearly independent for much of the time data were collected. Apparent linear dependence is restricted to late-time data, implying parameters cannot be uniquely estimated solely from late-time data. The parameter sensitivity curves obtained in both short- and long-pulse injection tests show a weak symmetry between two opposite-sign branches  associated with arrival and elution tracer breakthrough waves. Absolute values of sensitivity coefficients are largest when measured concentrations are changing most rapidly. Variation of sensitivity coefficients with time for retarding tracer $^{232}$U in test B3 are shown in Figure~\ref{fig:sen2}. These are also sufficiently larger than zero indicating that parameters, including $R_\mathrm{m}$ and $R_\mathrm{im}$, are estimable from breakthrough data.

Parameter estimation was performed using PEST \citep{doherty2010}. The optimal vector of model parameters $(\bm{\theta}_\mathrm{opt})$ was obtained by minimizing the sum of squared residuals,
\begin{equation}
\label{eqn:obj}
\Phi (\bm{\theta}) = \bm{e}(\bm{\theta})^T\bm{e}(\bm{\theta}),
\end{equation}
where $\bm{e} = \bm{c}_\mathrm{obs} - \bm{c}_\mathrm{cal}(\bm{\theta})$ is the vector of residuals. PEST uses the Levenberg-Marquardt nonlinear optimization algorithm \citep{marquardt1963}. Parameter estimates and multirate model fits to data are compared to those obtained using single- and double-porosity models. Parameter values obtained by inverting $^3$H and $^{22}$Na breakthrough data with the all three models are summarized in Table~\ref{tab:core-B}; parameters estimated from $^{232}$U data are in Table~\ref{tab:232U}. Because $t_\mathrm{inj}$ was not reported in the original study \citep{lucero1998}, it was estimated from data. The $\omega_D$ column also includes the mean $(\langle \omega_D \rangle$) and variance ($\sigma_{\Omega_D}^2$) of the Damk\"{o}hler-I number determined from the estimated values of $\hat{\mu}$ and $\hat{\sigma}$. The last row of Table~\ref{tab:core-B} shows estimated model parameters from simultaneous inversion of B4, B5, and B8 tracer-test breakthrough data. Parameter estimates are comparable to those from individual tests, even though the three tests were conducted with flow rates ranging over an order of magnitude (0.05, 0.1, and 0.5~ml/min). This indicates minimal model structural error with regard to simulating average pore-water velocity.

\begin{figure}[!hbtp]
  \centering
 \includegraphics[height=0.95\textheight]{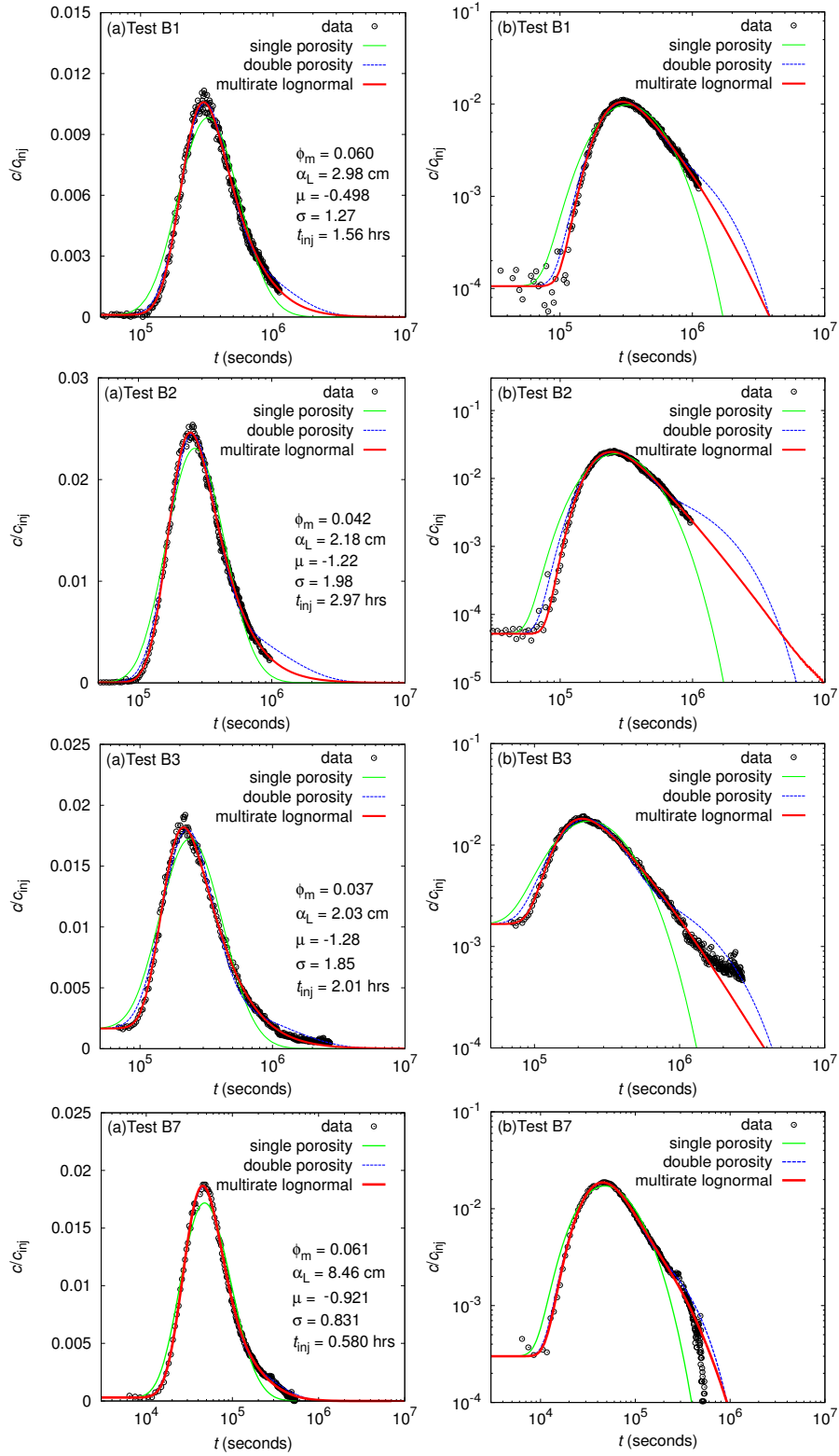}
  \caption{ Model fits to core B $^{22}\mathrm{Na}$ breakthrough data
    for short-pulse tests (B1, B2, B3, and B7) on (a) semi-log and (b)
    log-log scales.}
\label{fig:modelfits1}
\end{figure}

\begin{figure}[!hbtp]
  \centering
  \includegraphics[height=0.75\textheight]{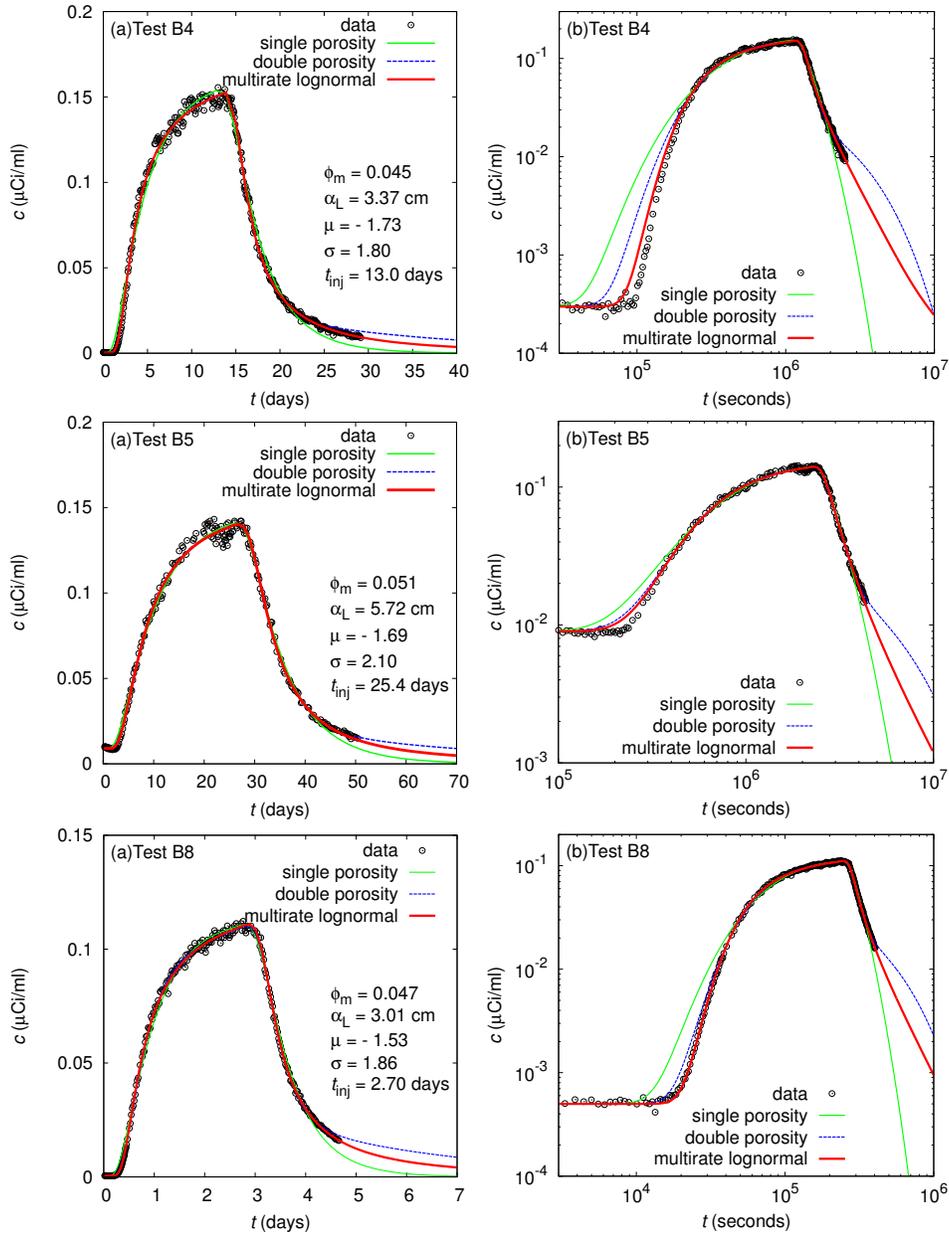} 
  \caption{Model fits to core B $^{22}\mathrm{Na}$ breakthrough data
    for long-pulse tests (B4, B5, and B8) on (a) linear and (b)
    log-log scales.}
\label{fig:modelfits2}
\end{figure}

\begin{figure}[!hbtp]
  \includegraphics[width=\textwidth]{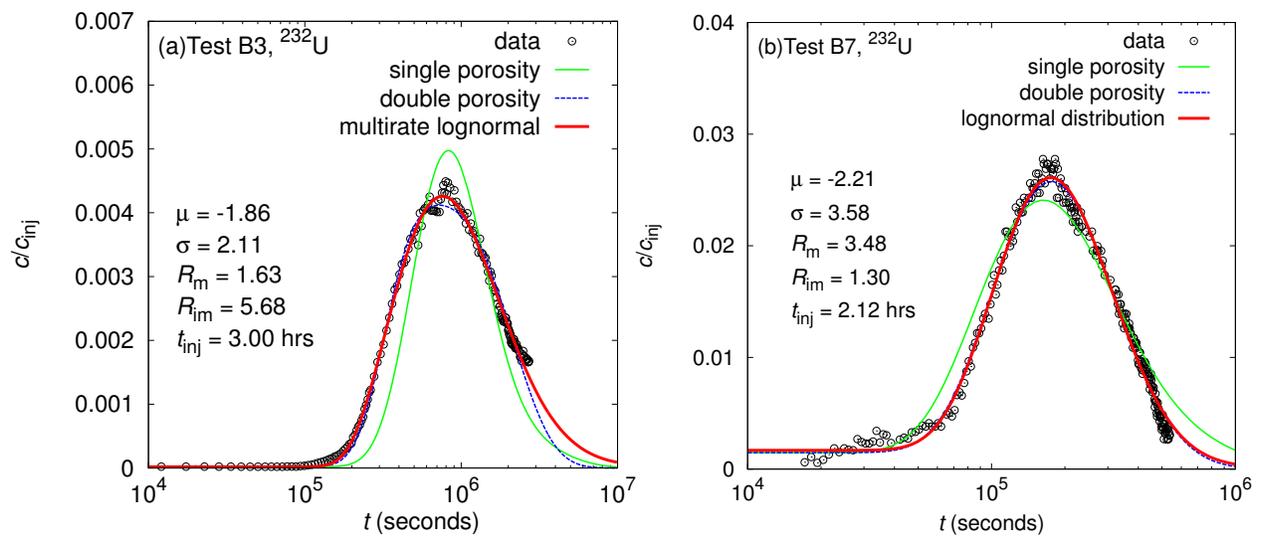} 
  \caption{Model fits to $^{232}\mathrm{U}$ breakthrough data from
    tests B3 and B7 on semi-log plots. Parameters shown in plots are
    for the multirate model.}
\label{fig:modelfits3}
\end{figure}

Model fits to data for parameter values listed in Table~\ref{tab:core-B} are shown in Figure~\ref{fig:modelfits1} (B1--B3, B7) and Figure~\ref{fig:modelfits2} (B4, B5, B8) for $^3$H and $^{22}$Na. Figures are in pairs of (a) linear or semi-log (concentration on linear scale) and (b) log-log plots, to illustrate how models match data over multiple time scales and over several concentration orders of magnitude. The two plotting scales are complementary because an apparently good model fit on a semi-log or linear plot may reveal a poor fit on log-log scale, and vice versa. Model-fit results for $^{232}$U data are shown in Figure~\ref{fig:modelfits3}. \citet{lucero1998} parameter estimates are comparable to those obtained here using single- and double-porosity models, but they did not estimate $t_\mathrm{inj}$.

Parameter estimation using the multirate model yielded improved model fits to breakthrough data compared to those obtained using single- and double-porosity models (see $R^2$ values in Table~\ref{tab:core-B}). Mobile-domain porosity values ($\phi_\mathrm{m}$) estimated with single- and double-porosity models were comparable (means of $0.069$ and $0.065$, respectively), but were appreciably larger than those obtained using the multirate model (mean of $0.045$). Dispersivity ($\alpha_L$) values were consistently largest for the single-porosity model (mean of $12.1$~cm) and smallest for the multirate model (mean of $3.76$~cm) for all tests. Table~\ref{tab:core-B} shows there is significantly more variability in $\alpha_L$ estimated using the single-porosity model than those obtained using the double-porosity and multirate models (standard deviations of 4.2~cm, 2.4~cm, and 2.3~cm, respectively). The Damk\"{o}hler-I numbers estimated with the double-porosity model appear closer (though not equal) to the geometric mean ($\langle \omega_D \rangle_g = e^\mu$) of the multirate model than to the mean ($\langle \omega_D \rangle = e^{\mu+\sigma^2/2}$). Results show absolute values of $\mu$ and $\sigma$ for the $^3$H tracer test (B1) are smaller than those obtained with the tracer $^{22}$Na. With exception of B7, the $^{22}$Na tests yielded consistent values of $\mu$ and $\sigma$ with $|\mu| > 1.0$ and $\sigma \approx 1.9$. Those obtained from $^{232}$U data (Table~\ref{tab:232U}) are significantly different.

For the non-conservative tracer $^{232}$U, $\phi_\mathrm{m}$ and $\alpha_L$ were estimated with the $^{22}$Na tracer from the same experiment, because these parameters are intrinsic transport medium properties. Estimated retardation factors from tests B3 and B7 are listed in Table~\ref{tab:232U}. For test B3, fitting the multirate and double-porosity models to data yields $R_\mathrm{m}$ values appreciably smaller than the value obtained with single-porosity model. This is because retardation is distributed between the mobile and immobile domains in the former two models.  It is surprising to find the multirate model $R_\mathrm{im}$ in test B3 is significantly larger than the double-porosity model $R_\mathrm{im}$. Intuitively, one would expect results similar to those obtained from test B7, because delayed breakthrough is partly due to matrix mass transfer and partly due to solid-phase sorption. In addition, the retardation factors, $R_\mathrm{m}$ and $R_\mathrm{im}$, estimated with the double-porosity and multirate models showed significant differences between test B3 and B7. These two results may be attributable to interplay between multirate mass-transfer and nonlinear sorption kinetics, where retardation is concentration dependent. The models all assume linear instantaneous sorption,  variability in retardation factors between tests B3 and B7 may be an artifact of inherent model deficiency to account for nonlinear sorption kinetics. $^{232}$U column tests tests B3, B6 (not discussed here) and B7 were performed serially on the same core. B3 had the lowest initial relative $^{232}$U concentration with $c_0/c_\mathrm{inj} \simeq 2 \times 10^{-5}$, while for B7 $c_0/c_\mathrm{inj} \simeq 10^{-3}$. B7 was performed after the core had already been conditioned with $^{232}$U from the previous two tests. These initial concentration differences are expected to affect the estimated retardation factors in the presence of nonlinear sorption kinetics.

\subsection{Predictive Analysis}
All models approximate a complex reality, and the discrepancy between reality and mathematical models is commonly referred to as model structural error. It is a measure of model deficiencies that lead to prediction errors even when the models are supplied with optimal input parameters. Structural error cannot be attributed to measurement errors inherent in observations \citep{doherty2010b} and typically decreases as models become more realistic with increased understanding of underlying causal mechanisms of processes. A measure of structural error would thus provide an objective criterion for model selection.

\begin{figure}[!htbp]
  \centering
  \includegraphics[height=0.8\textheight]{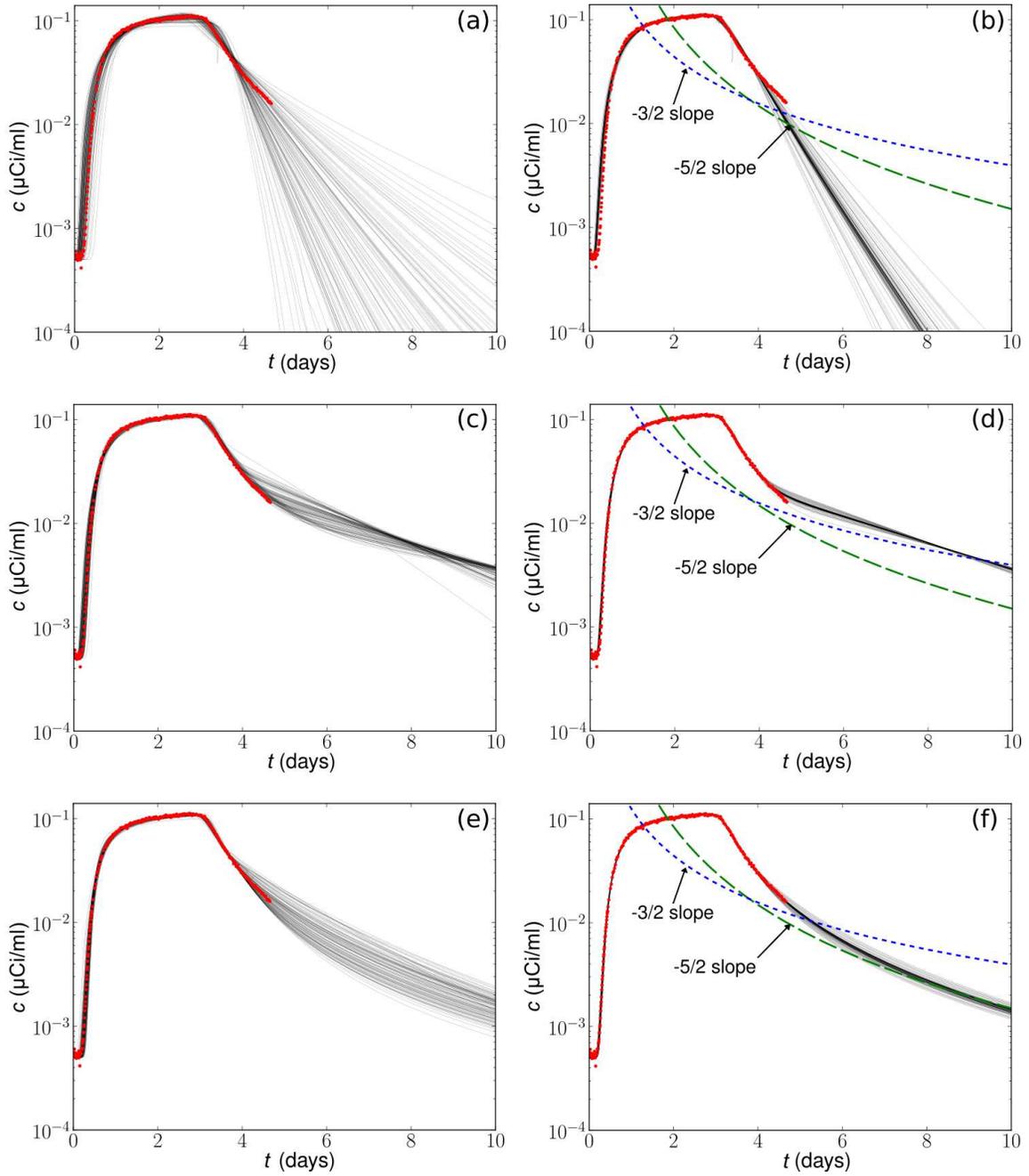}
  \caption{Model prediction uncertainty evaluated using posterior
    Monte Carlo analysis on B8 data with the (a,b) single-porosity,
    (c,d) double-porosity and (e,f) the multirate models (left and
    right columns represent before and after re-calibration,
    respectively). Double porosity model predictions approach $-3/2$
    slope while the multirate model predictions approach $-5/2$
    slope.} \label{fig:pred-anal}
\end{figure}

Predictive uncertainty analysis presented here is used to demonstrate the structural deficiency of the single- and double-porosity models, and how this deficiency leads to increased model prediction error. The analysis was undertaken with PEST for test B8. Details for conducting a PEST predictive uncertainty analysis can be found elsewhere \citep{james2009,tonkin2009,tonkin2007,gallagher2007}. Using parameter values at optimality (Table~\ref{tab:core-B}) and the associated covariance matrix, 500 random parameter sets were generated and projected onto the Jacobian matrix null space. No clear null space was found  from the singular value decomposition of the Jacobian matrix, therefore we assumed the null space to be a single dimension in these low-dimensional ($\le 6$) models. Model predictions computed beyond the last observation based on the 500 parameter sets generated in this manner are shown in the left column of Figure~\ref{fig:pred-anal} for (a) single-porosity, (c) double-porosity, and (e) multirate models. They show significant model prediction uncertainty for the single-porosity model, and only moderate uncertainty for the other two models. Using these parameter sets projected onto the null space as initial guesses, further minimization of $\Phi$ was undertaken, using the Jacobian matrix associated with the calibrated state. Using the value of $\Phi$ at optimality $(\Phi_\mathrm{opt})$, the 500 null-space-projected parameter sets were processed with PEST to minimize the objective function such that $\Phi \leq 2 \Phi_\mathrm{opt}$. Predictions associated with the re-calibrated parameter sets are shown in the second column of Figure~\ref{fig:pred-anal} for (b) single-porosity, (d) double-porosity, and (f) multirate models. As would be expected, post re-calibration model predictions for all three models show a marked decrease in model prediction uncertainty from the pre-calibration predictions. The late-time $-3/2$ and $-5/2$ slope lines are included, which are diagnostic of double-porosity and multirate models \citep{haggerty2000}. Clearly, the model behavior projected beyond the time of the last observation follows the $-3/2$ slope for the dual-porosity model, and the $-5/2$ slope for the multirate mass transfer model.

\begin{figure}[!htbp]
  \centering
  \includegraphics[width=0.9\textwidth]{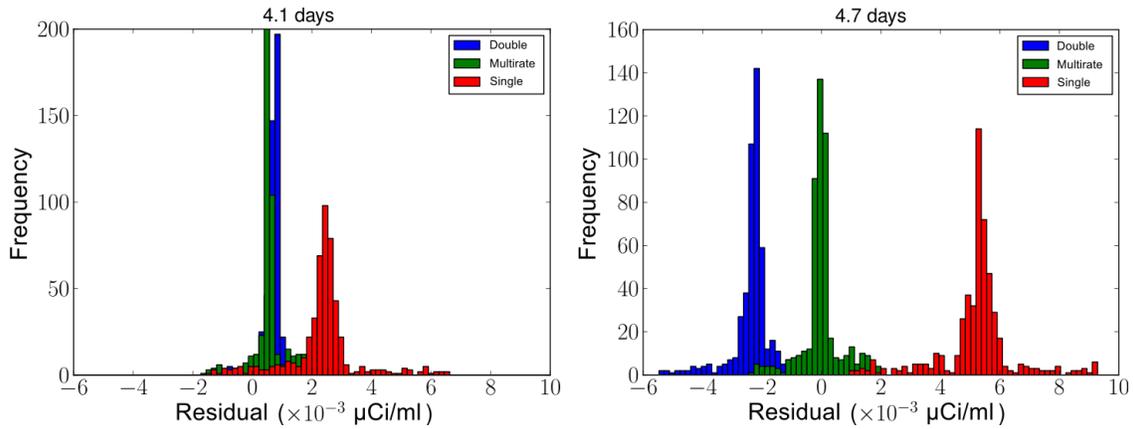} 
  \caption{Residual histograms computed at (a) $t=4.1$ days and (b)
    $t=4.7$ days with re-calibrated model
    runs.} \label{fig:pred-residuals}
\end{figure}

Re-calibration single-porosity model projections show significant underestimation of late-time observations. Dual-porosity model predictions are skewed toward overestimating the late-time observations. Multirate model projections are uniformly centered about the data and are consistent with the observed trend of the elution curve. Figure~\ref{fig:pred-residuals} shows histograms of residuals associated with the three models plotted at (a) $t=4.1$ and (b) $t=4.7$ days. Whereas the residuals computed at $t=4.7$ days with the multirate model have zero bias, those of the double- and single-porosity models show clear bias to negative (concentration overestimation) and positive (underestimation) values. Only the multirate model shows minimal bias about the observed late-time data, even though its ensemble of predictions has comparable spread (variance) to those of the double-porosity model beyond the last observation. The residual bias signifies model structural error associated with single- and dual-porosity models. Comparing results in Figure~\ref{fig:pred-residuals} (a) and (b) shows residual bias and single- and double-porosity model structural error increase with time, while bias for the multirate model does not show appreciable change. At time $t=4.1$ days, the dual-porosity model residuals have zero mean and are nearly coincident with the multirate model. However, at $t=4.7$ days there is a growth in double-porosity model prediction bias. Prediction error due to model structural error increases with time.

\begin{figure}[!htbp]
  \centering
  \includegraphics[width=0.8\textwidth]{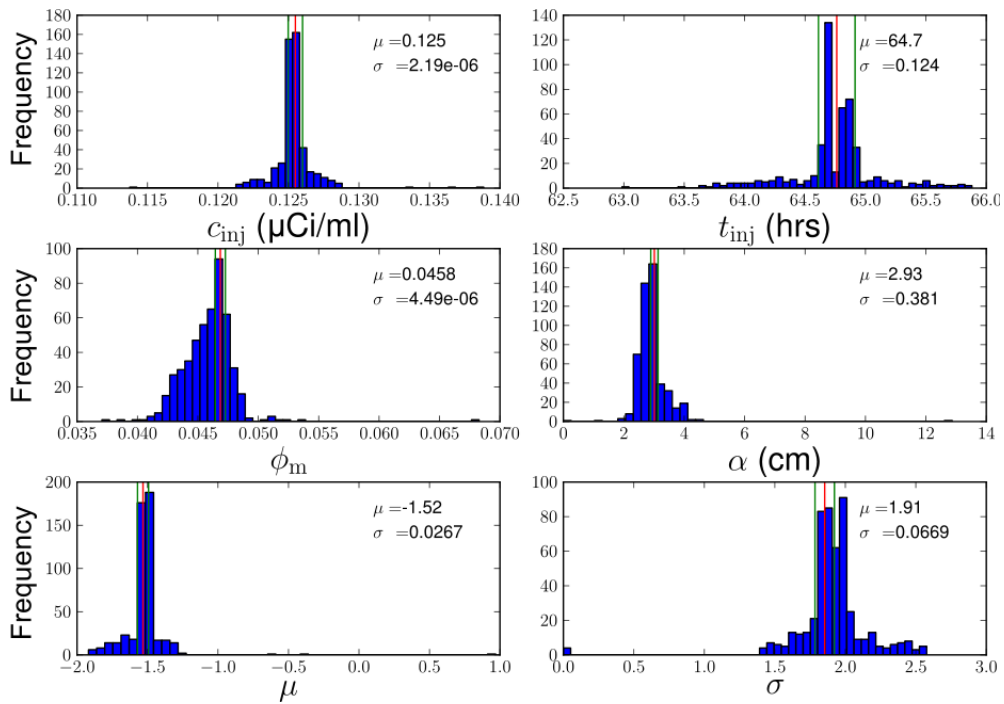}
  \caption{Parameter histograms after re-calibration with PEST
    posterior Monte Carlo analysis for test B8. Red line indicate
    PEST-estimated optimal parameter values and green lines are 
    PEST-estimated 95\% confidence intervals.} \label{fig:params-hist}
\end{figure}

\begin{figure}[!htbp]
  \centering
  \includegraphics[width=\textwidth]{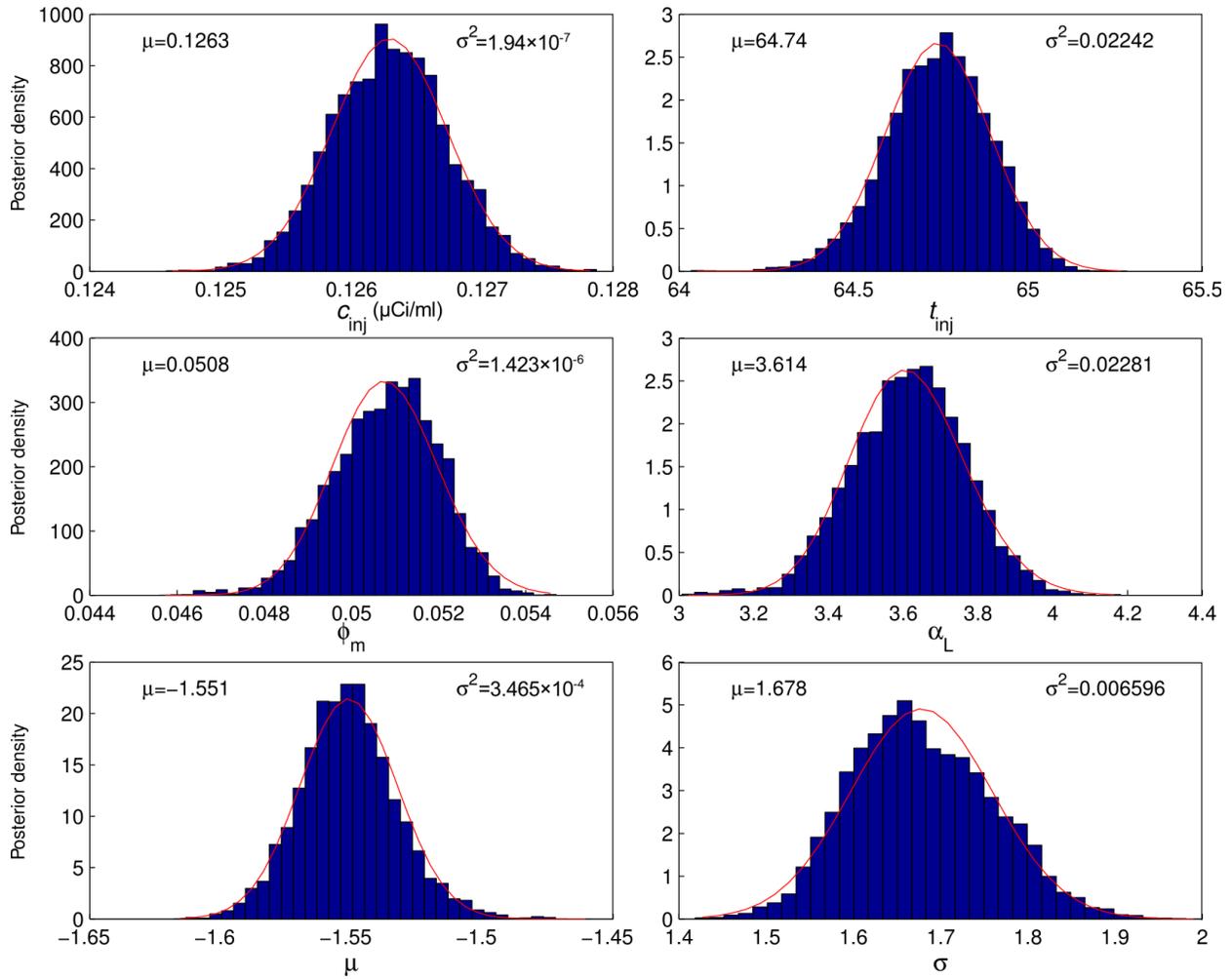}
  \caption{Posterior model parameter distributions estimated with
    DREAM for test B8. Red curves are normal distributions
    corresponding to mean and variance computed from
    data.} \label{fig:pred-params-dream}
\end{figure}

Figure~\ref{fig:params-hist} shows histograms of 500 calibrated multirate model parameter sets obtained from the posterior null-space Monte Carlo analysis described above. These distributions provide a measure of parameter estimation uncertainty. However, as indicated  by \citet{keating2010}, parameter sets obtained using null-space Monte Carlo analysis do not necessarily constitute a sample of the posterior density function of the parameters in the strict Bayesian sense. This is especially true with low-dimensional models (at most 6 parameters for the present case) for which a proper null space may not exist. This can be seen by comparing the posterior distribution obtained with the null-space Monte Carlo analysis with those obtained to a formal Bayesian approach using the DiffeRential Evolution Adaptive Metropolis (DREAM) algorithm \citep{vrugt2008,vrugt2009a,vrugt2009b}. For the problem considered here with 6 parameters to be estimated from log-transformed concentrations, DREAM ran 6 different Markov chains, and after a burn-in period of about 35,000 model runs per chain, we obtained the parameter posterior distributions shown in Figure~\ref{fig:pred-params-dream}. DREAM required 300,000 total model runs. Clearly, the computational demands of formal Bayesian analysis with DREAM can be prohibitively high \citep{keating2010}. The parameter posterior distributions shown in Figure~\ref{fig:pred-params-dream} show the final 10,000 model runs. Normal distributions are included in the figure for comparison. The results show that posterior distributions obtained with DREAM have smaller variances and are more Gaussian than those obtained with the PEST posterior null-space Monte Carlo analysis. Whilst PEST results indicate greater variability in estimated parameter values that calibrate the model, DREAM results  indicate that parameter estimation uncertainty is actually smaller. The low-dimensionality of the parameter space leads to an overestimation of parameter estimation uncertainty using null-space Monte-Carlo analysis. Thus, PEST-based parameter estimation uncertainty, obtained with null-space Monte Carlo analysis for a significantly lower computational cost, may be viewed as the upper bound of the true uncertainty computed with DREAM, for cases like the low-dimensional models used here.

\subsection{Statistical Model Selection}
For a given number of obervations, as models become more realistic, the increase in model complexity and the number of parameters leads to increased parameter estimation uncertainty because the number of observations available per estimated parameter decreases. In the present case, model complexity and the number of parameters increase from the single-porosity to the multirate model, but the respective model parameters are estimated with the same number of observations. Hence, statistical criteria that account for decreased information content due to increased model complexity may be used to augment  model selection based on structural error evaluation. The corrected Akaike Information Criterion, $\mathrm{AIC_c}$ \citep{hurvich1997, anderson1999, poeter2005} is used here for this purpose
\begin{equation}
  \label{eqn:aicc}
  \mathrm{AIC_c} = 2n\left[\log(\sigma_e) + \frac{k}{n-k-1}\right],
\end{equation}
where $n$ is the number of observations, $k$ is the number of estimated parameters, and $\sigma_e$ is the standard deviation of residuals at optimality. The first term typically decreases as model complexity increases, representing improved model fit to data, while the second penalty term increases. Because $\mathrm{AIC_c}$ is a relative measure, it is preferable to use differentials of $\mathrm{AIC_c}$ \citep{posada2004}, denoted $\Delta \mathrm{AIC_c}$, over all the three models under consideration. For the $i^\mathrm{th}$ model, $\Delta \mathrm{AIC}_{\mathrm{c},i} = \mathrm{AIC}_{\mathrm{c},i} - \min \mathrm{AIC_c}$, where $\min \mathrm{AIC_c}$ is the smallest $\mathrm{AIC_c}$ value among all models for this dataset. The $\mathrm{AIC_c}$ are computed using PEST and $\Delta \mathrm{AIC_c}$ are listed in Table \ref{tab:core-B}. The minimum $\mathrm{AIC_c}$ corresponds to the multirate model, except in test B5, where the it corresponds to the double-porosity model. Clearly the relative $\mathrm{AIC_c}$ values confirm the results of predictive analysis that the multirate model is better suited than the other two models to describing transport in the Culebra Dolomite core.

\begin{figure}[!htbp]
  \centering
  \includegraphics[height=0.35\textheight]{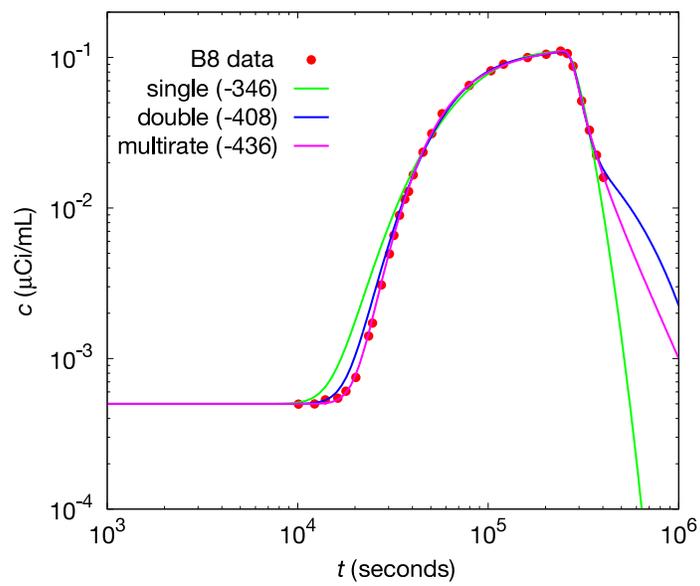}
  \caption{\label{fig:reduceddataB8}Model fits to test B8 data where only 30 data points were used in the optimization. The $\mathrm{AIC_c}$ for each model is included in parenthesis.}
\end{figure}

For time series data with high autocorrelation, the penalty for model complexity is vanishingly small when $n \ggg k$ and the $\mathrm{AIC_c}$ reduces to a ranking of the  models by residual variance. This is only a problem however, when the increased number of observations does not singinficantly increase the information content of the observation about the estimated parameters. Hence, a separate optimization with PEST using only 30 of the original 269 data in test B8 to determine whether the ranking of the three models with the $\mathrm{AIC_c}$ would change appreciably. The resulting model fits are shown in Figure \ref{fig:reduceddataB8}. Basically the same results were obtained, with the multirate model outperforming the other two models. This is because the estimation variance is always smallest for the multirate model, and artificially reducing $n$ only has a modest effect on the final outcome. It should also be noted that a large $n$ allows one to better capture the variability in the data due to random measurement error, which are assumed to be Gaussian in minimization of the sum of squared residuals. Further, the number of parameters to be estimated increased only by 2 from the single-porosity to the multirate model, whereas the estimation variance changes by a factor of about 2 ($7.6\times 10^{-6}$ to $3.2\times 10^{-7}$).

\begin{figure}[!htbp]
  \begin{center}
    \includegraphics[height=0.5\textheight]{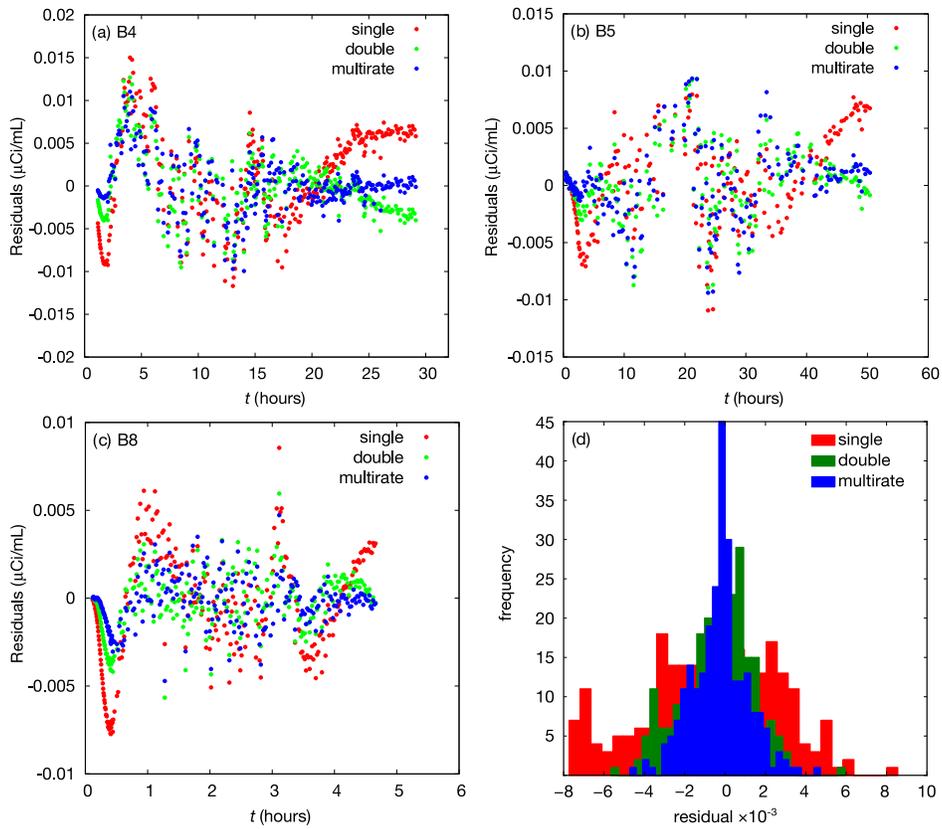}
  \end{center}
  \caption{\label{fig:residuals}Temporal residuals of tests B4, B5 and B8 and the histogram of the test B8 residuals.}
\end{figure}

The temporal structure of the residuals was examined to determine whether they show strong temporal autocorrelation. They are plotted in Figure \ref{fig:residuals}. It can be seen in the Figure that moderate autocorrelation is limited to very early-time. Additionally, in this early-time period, it can be seen that only the single-porosity model residuals show appreciable temporal autocorrelation, which decreases as one moves to the multirate model. The computed responses of the single-porosity model show strong departure from observed behavior. As can be seen in Figure \ref{fig:residuals}, the residuals obtained with the multirate model for the long tests (B4, B5 \& B8) show only moderate temporal autocorrelation (at early-time) and are mostly randomly distributed about zero. It should also be noted that the statistical rigor of DREAM does not depend on the distribution of the residuals but on the sampling of the parameter space for parameters that minimize the sum of squared residuals.

\section{Discussion and Conclusions}
We reanalyzed core-scale $^3$H and $^{22}$Na breakthrough data from experiments conducted by \citet{lucero1998} on a Culebra Dolomite core using the single-porosity, double-porosity, and the multirate model of \citep{haggerty1995, haggerty1998} on a semi-infinite domain to determine which of the models best describes the observed breakthrough behavior. Previous analysis of these data by \citet{lucero1998} had suggested that the single-porosity model was sufficient to describe core-scale Culebra transport, a finding that was at odds with findings based on field-scale tests conducted in the Culebra Dolomite formation \citep{meigs2000, mckenna2001}. In the results presented herein, the multirate model yielded better model fits to the data and parameter values that differed significantly from those obtained with the single- and double-porosity models. The mobile-domain porosity and dispersivity values obtained with the multirate model were consistently lower than those obtained with the other two models because solute dispersion in the core is also accounted for by porosity variability encapsulated in the distribution parameters of the mobile/immobile domain mass-transfer coefficient. The smaller dispersivity obtained with the multirate model is more plausible than those obtained with the other models, considering the length scale of the experiments.

Model-prediction uncertainty was evaluated using breakthrough data from test B8 and post-calibration null-space Monte Carlo analysis as implemented with PEST. The prediction uncertainty analysis revealed the presence of model structural error in the single- and double-porosity models as demonstrated by significant bias in the residuals of model predictions made with these models with optimal parameter values. The residual bias increased with time over the span of the elution curve where breakthrough data are available, showing increased departure of model predictions from the observed trend ($-5/2$ slope line) of breakthrough data. The parameters associated with the null-space Monte Carlo predictive analysis may be viewed as samples from the posterior parameter distributions and were used to evaluate parameter estimation uncertainty. The posterior distributions estimated using null-space Monte Carlo analysis were compared to those obtained with the more rigorous Bayesian analysis in the DREAM algorithm. The comparison suggests that measures of parameter estimation uncertainty obtained with null-space Monte Carlo may be treated as upper bounds of the true posterior distributions, particularly for low dimensional models where a true null space may not exist.

The analysis presented herein clearly shows the residual bias associated with the single- and double-porosity models increases with time indicating increasing systematic departure of predicted from observed behavior due to the inherent structural deficiencies of these models. The multirate model residuals, however, maintain minimal bias with time, indicating low model structural error. Although the predictions with the double-porosity and multirate models beyond the last observation have comparable variance, only the residuals of the multirate model have zero bias. These results show that the multirate model is the most appropriate of the three models for describing solute breakthrough behavior in Culebra core even though the three models yield parameters with comparable variances of posterior distributions. This finding was further confirmed using statistical model selection using the differential $\mathrm{AIC_c}$ where the $\mathrm{AIC_c}$ value was typically smallest for the multirate model. The one test where the double-porosity model yielded the smallest differential $\mathrm{AIC_c}$ value, the value associated with the multirate model was only marginally larger (0.5\%). More elution data would be needed to resolve this minor departure from the norm given that the two models predict disparate long-term tailing behaviors.

\section*{Acknowledgements}
Sandia National Laboratories is a multi-program laboratory managed and operated by Sandia Corporation, a wholly owned subsidiary of Lockheed Martin Corporation, for the U.S. Department of Energy's National Nuclear Security Administration under contract DE-AC04-94AL85000.  This research is funded by WIPP programs administrated by the Office of Environmental Management (EM) of the US Department of Energy.

\bibliographystyle{abbrvnat} 
\bibliography{transport}

\begin{thebibliography}{35}
\providecommand{\natexlab}[1]{#1}
\providecommand{\url}[1]{\texttt{#1}}
\expandafter\ifx\csname urlstyle\endcsname\relax
  \providecommand{\doi}[1]{doi: #1}\else
  \providecommand{\doi}{doi: \begingroup \urlstyle{rm}\Url}\fi

\bibitem[Anderson and Burnham(1999)]{anderson1999}
D.~R. Anderson and K.~P. Burnham.
\newblock Understanding information criteria for selection among
  capture-recapture or ring recovery models.
\newblock \emph{Bird Study}, 46\penalty0 (supplement):\penalty0 S14--21, 1999.

\bibitem[Buchan et~al.(1993)Buchan, Grewal, and Robson]{buchan1993}
G.~D. Buchan, K.~S. Grewal, and A.~B. Robson.
\newblock Improved models of particle-size distribution: An illustration of
  model comparison techniques.
\newblock \emph{Soil Science Society of America Journal}, 57\penalty0
  (4):\penalty0 901--908, 1993.

\bibitem[{de Hoog} et~al.(1982){de Hoog}, Knight, and Stokes]{dehoog1982}
F.~R. {de Hoog}, J.~H. Knight, and A.~N. Stokes.
\newblock An improved method for numerical inversion of {L}aplace transforms.
\newblock \emph{{SIAM} Journal of Scientific and Statistical Computing},
  3\penalty0 (3):\penalty0 357--366, 1982.

\bibitem[Doherty(2010)]{doherty2010}
J.~Doherty.
\newblock \emph{{PEST User-Manual}: Model-independent parameter estimation}.
\newblock Watermark Numerical Computing, Australia, fifth edition, 2010.

\bibitem[Doherty and Welter(2010)]{doherty2010b}
J.~Doherty and D.~E. Welter.
\newblock A short exploration of structural noise.
\newblock \emph{Water Resources Research}, 46\penalty0 (5):\penalty0 W05525,
  2010.

\bibitem[Gallagher and Doherty(2007)]{gallagher2007}
M.~Gallagher and J.~E. Doherty.
\newblock Predictive error analysis for a water resource management model.
\newblock \emph{Journal of Hydrology}, 334\penalty0 (3-4):\penalty0 513--533,
  2007.

\bibitem[Gamerdinger et~al.(1990)Gamerdinger, Wagenet, and van
  Genuchten]{gamerdinger1990}
A.~P. Gamerdinger, R.~J. Wagenet, and M.~T. van Genuchten.
\newblock Application of two-site/two-region models for studying simultaneous
  nonequilibrium transport and degradation of pesticides.
\newblock \emph{Soil Science Society of America Journal}, 54\penalty0
  (4):\penalty0 957--963, 1990.

\bibitem[Haggerty and Gorelick(1995)]{haggerty1995}
R.~Haggerty and S.~M. Gorelick.
\newblock Multiple-rate mass-transfer for modeling diffusion and
  surface-reactions in media with pore-scale heterogeneity.
\newblock \emph{Water Resources Research}, 31\penalty0 (10):\penalty0
  2383--2400, 1995.

\bibitem[Haggerty and Gorelick(1998)]{haggerty1998}
R.~Haggerty and S.~M. Gorelick.
\newblock Modeling mass transfer processes in soil columns with pore-scale
  heterogeneity.
\newblock \emph{Soil Science Society of America Journal}, 62\penalty0
  (1):\penalty0 62--74, 1998.

\bibitem[Haggerty et~al.(2000)Haggerty, McKenna, and Meigs]{haggerty2000}
R.~Haggerty, S.~A. McKenna, and L.~C. Meigs.
\newblock On the late-time behavior of tracer test breakthrough curves.
\newblock \emph{Water Resources Research}, 36\penalty0 (12):\penalty0
  3467--3479, 2000.

\bibitem[Haggerty et~al.(2001)Haggerty, Fleming, Meigs, and
  McKenna]{haggerty2001}
R.~Haggerty, S.~W. Fleming, L.~C. Meigs, and S.~A. McKenna.
\newblock Tracer tests in a fractured dolomite: 2. analysis of mass transfer in
  single-well injection-withdrawal test.
\newblock \emph{Water Resources Research}, 37\penalty0 (5):\penalty0
  1129--1142, 2001.

\bibitem[Hoeksema and Kitanidis(1985)]{hoeksema1985}
R.~J. Hoeksema and P.~K. Kitanidis.
\newblock Analysis of the spatial structure of properties of selected aquifers.
\newblock \emph{Water Resources Research}, 21\penalty0 (4):\penalty0 563--572,
  1985.

\bibitem[Holt(1997)]{holt1997}
R.~M. Holt.
\newblock Conceptual model for transport processes in the {Culebra {D}olomite
  Member, Rustler Formation}.
\newblock Technical Report SAND97--0194, Sandia National Laboratories, 1997.

\bibitem[Hurvich and Tsai(1997)]{hurvich1997}
C.~M. Hurvich and C.-L. Tsai.
\newblock Selection of a multistep linear predictor for short time series.
\newblock \emph{Statistica Sinica}, 7\penalty0 (2):\penalty0 395--406, 1997.

\bibitem[Huyakorn et~al.(1983)Huyakorn, Lester, and Mercer]{huyakorn1983}
P.~S. Huyakorn, B.~H. Lester, and J.~W. Mercer.
\newblock An efficient finite element technique for modeling transport in
  fractured porous media: 1. single species transport.
\newblock \emph{Water Resources Research}, 19\penalty0 (3):\penalty0 841--854,
  1983.

\bibitem[James et~al.(2009)James, Doherty, and Eddebbarh]{james2009}
S.~C. James, J.~D. Doherty, and A.-A. Eddebbarh.
\newblock Practical postcalibration uncertainty analysis: {Yucca Mountain,
  Nevada}.
\newblock \emph{Ground Water}, 47\penalty0 (6):\penalty0 851--869, 2009.

\bibitem[Keating et~al.(2010)Keating, Doherty, Vrugt, and Kang]{keating2010}
E.~H. Keating, J.~Doherty, J.~A. Vrugt, and Q.~J. Kang.
\newblock Optimization and uncertainty assessment of strongly nonlinear
  groundwater models with high parameter dimensionality.
\newblock \emph{Water Resources Research}, 46:\penalty0 W10517, 2010.

\bibitem[Lucero et~al.(1998)Lucero, Brown, and Heath]{lucero1998}
D.~A. Lucero, G.~O. Brown, and C.~H. Heath.
\newblock Laboratory column experiments for radionuclide adsorption studies of
  the {C}ulebra {D}olomite {M}ember of the {R}ustler {F}ormation.
\newblock Technical Report SAND97--1763, Sandia National Laboratories,
  Albuquerque, NM, 1998.

\bibitem[Marquardt(1963)]{marquardt1963}
D.~Marquardt.
\newblock An algorithm for least-squares estimation of nonlinear parameters.
\newblock \emph{Journal of the Society for Industrial and Applied Mathematics},
  11\penalty0 (2):\penalty0 431--441, 1963.

\bibitem[McKenna et~al.(2001)McKenna, Meigs, and Haggerty]{mckenna2001}
S.~A. McKenna, L.~C. Meigs, and R.~Haggerty.
\newblock Tracer tests in a fractured dolomite 3. double-porosity,
  multiple-rate mass transfer processes in convergent flow tracer tests.
\newblock \emph{Water Resources Research}, 37\penalty0 (5):\penalty0
  1143--1154, 2001.

\bibitem[Meigs and Beauheim(2001)]{meigs01}
L.~C. Meigs and R.~L. Beauheim.
\newblock Tracer tests in a fractured dolomite 1. experimental design and
  observed tracer recoveries.
\newblock \emph{Water Resources Research}, 37\penalty0 (5):\penalty0
  1113--1128, 2001.

\bibitem[Meigs et~al.(2000)Meigs, Beauheim, and Jones]{meigs2000}
L.~C. Meigs, L.~C. amd~Chambers, R.~L. Beauheim, and T.~L. Jones.
\newblock Interpretations of tracer tests performed in the {C}ulebra {D}olomite
  at the {W}aste {I}solation {P}ilot {P}lant site.
\newblock Technical Report SAND97--3109, Sandia National Laboratories,
  Albuquerque, NM, 2000.

\bibitem[Neuman(1982)]{neuman1982}
S.~P. Neuman.
\newblock Statistical characterization of aquifer heterogeneities: An overview.
\newblock \emph{Special Paper - Geological Society of America}, 189:\penalty0
  81--102, 1982.

\bibitem[\"{O}zi\c{s}ik and Orlande(2000)]{ozisik2000}
M.~N. \"{O}zi\c{s}ik and H.~R.~B. Orlande.
\newblock \emph{Inverse heat transfer: Fundamentals and applications}.
\newblock Taylor \& Francis, New York, 2000.

\bibitem[Poeter and Anderson(2005)]{poeter2005}
E.~Poeter and D.~R. Anderson.
\newblock Multimodel ranking and inference in ground water modeling.
\newblock \emph{Ground Water}, 43\penalty0 (4):\penalty0 597--605, 2005.

\bibitem[Posada and Buckley(2004)]{posada2004}
D.~Posada and T.~R. Buckley.
\newblock {Model selection and model averaging in phylogenetics: Advantages of
  Akaike Information Criterion and Bayesian approaches over likelihood ratio
  tests}.
\newblock \emph{Systematic Biology}, 53\penalty0 (5):\penalty0 793--808, 2004.

\bibitem[Schumer et~al.(2003)Schumer, Benson, Meerschaertm, and
  Baeumer]{schumer2003}
R.~Schumer, D.~A. Benson, M.~M. Meerschaertm, and B.~Baeumer.
\newblock Fractal mobile/immobile solute transport.
\newblock \emph{Water Resources Research}, 39\penalty0 (10):\penalty0 1296,
  2003.

\bibitem[Sun and Buscheck(2003)]{sun2003}
Y.~W. Sun and T.~A. Buscheck.
\newblock Analytical solutions for reactive transport of {N}-member
  radionuclide chains in a single fracture.
\newblock \emph{Journal of Contaminant Hydrology}, 62--63:\penalty0 695--712,
  2003.

\bibitem[Tonkin and Doherty(2009)]{tonkin2009}
M.~J. Tonkin and J.~Doherty.
\newblock Calibration-constrained {M}onte {C}arlo analysis of highly
  parameterized models using subspace techniques.
\newblock \emph{Water Resources Research}, 45:\penalty0 W00B10, 2009.

\bibitem[Tonkin et~al.(2007)Tonkin, Doherty, and Moore]{tonkin2007}
M.~J. Tonkin, J.~E. Doherty, and C.~Moore.
\newblock Efficient nonlinear predictive error variance for highly
  parameterized models.
\newblock \emph{Water Resources Research}, 43\penalty0 (7):\penalty0 W07429,
  2007.

\bibitem[van Genuchten and Wagenet(1989)]{genuchten1989}
M.~T. van Genuchten and R.~J. Wagenet.
\newblock Two-site/two-region models for pesticide transport and degradation:
  theoretical development and analytical solutions.
\newblock \emph{Soil Science Society of America Journal}, 53\penalty0
  (5):\penalty0 1303--1310, 1989.

\bibitem[Vrugt et~al.(2008)Vrugt, Stauffer, Wohling, Robinson, and
  Vesselinov]{vrugt2008}
J.~A. Vrugt, P.~H. Stauffer, T.~Wohling, B.~A. Robinson, and V.~V. Vesselinov.
\newblock Inverse modeling of subsurface flow and transport properties: A
  review with new developments.
\newblock \emph{Vadose Zone Journal}, 7\penalty0 (2):\penalty0 843--864, 2008.

\bibitem[Vrugt et~al.(2009{\natexlab{a}})Vrugt, Robinson, and
  Hyman]{vrugt2009a}
J.~A. Vrugt, B.~A. Robinson, and J.~M. Hyman.
\newblock Self-adaptive multimethod search for global optimization in
  real-parameter spaces.
\newblock \emph{IEEE Transactions on Evolutionary Computation}, 13\penalty0
  (2):\penalty0 243--259, 2009{\natexlab{a}}.

\bibitem[Vrugt et~al.(2009{\natexlab{b}})Vrugt, ter Braak, Diks, Robinson,
  Hyman, and Higdon]{vrugt2009b}
J.~A. Vrugt, C.~J.~F. ter Braak, C.~G.~H. Diks, B.~A. Robinson, J.~M. Hyman,
  and D.~Higdon.
\newblock Accelerating {M}arkov chain {M}onte {C}arlo simulation by
  differential evolution with self-adaptive randomized subspace sampling.
\newblock \emph{International Journal of Nonlinear Sciences and Numerical
  Simulation}, 10\penalty0 (3):\penalty0 273--290, 2009{\natexlab{b}}.

\bibitem[Zheng et~al.(2010)Zheng, Bianchi, and Gorelick]{zheng2010}
C.~Zheng, M.~Bianchi, and S.~M. Gorelick.
\newblock Lessons learned from 25 years of research at the {MADE} site.
\newblock \emph{Ground Water}, 49\penalty0 (5):\penalty0 649--662, 2010.

\end{thebibliography}

%%\end{article}

%%\end{linenumbers}

% Figure 1

% Figure 2

% Figure 3

%Figure 4

% Figure 5

% Figure 6

% Figure 7

% Figure 8

% Figure 9

% Figure 10

% Figure 11

% Figure 12

% Figure 13

% Table 1
\begin{table}[ht]
%%\small
\caption{PEST-Estimated Parameters Using Conservative Tracer Breakthrough Data}
\label{tab:core-B}
\begin{tabular}{|l|c|l|c|c|c|c|c|c|c|c|}
\hline
 & $Q_x$ & & & $\alpha_L$ & & & & $t_\mathrm{inj}$ &\\
 Test & {\tiny(ml/min)} & Model & $\phi_\mathrm{m}$ & (cm) & $\omega_{D}$ & $\mu$ & $\sigma$ & (hours)  & $R^2$ & $\Delta \mathrm{AIC_c}$\\
\hline
  & & Single & 0.081 & 7.99 &  &  & & 1.33 & 0.982 & 902\\
B1 $^{3}\mathrm{H}$ & 0.1 & Double & 0.073 & 4.83 & 0.735 &  & & 1.69 & 0.997 & 215\\
  & & Multirate & 0.060 & 2.98 & (1.36, 7.43) & $-0.498$ & 1.28 & 1.56 & 0.998 & 0\\
%\hline
  & & Single & 0.065 & 7.30 & &  & & 2.42 & 0.986 & 974\\
B2 $^{22}\mathrm{Na}$  & 0.1 & Double & 0.061 & 5.14 & 0.538 & & & 3.48 & 0.997 & 476\\
  & &  Multirate & 0.042 & 2.18 & (2.10, 217.9) & $-1.22$ & 1.98 & 2.97 & 0.999 & 0\\
%\hline
  & & Single & 0.062 & 8.55 & & & & 1.70 & 0.989 & 881\\
B3 $^{22}\mathrm{Na}$  & 0.1 & Double & 0.058 & 6.04 & 0.395 & & & 2.17 & 0.996 & 474\\
  & &  Multirate & 0.037 & 2.03 & (1.54, 70.3) & $-1.28$ & 1.85 & 2.01 & 0.999 & 0\\
%\hline
  & & Single & 0.065 & 17.5 & & & & 310.3 & 0.995 & 466\\
B4 $^{22}\mathrm{Na}$  & 0.1 & Double & 0.062 & 9.38 & 0.209 & & & 312.4 & 0.998 & 117\\
  & &  Multirate & 0.045 & 3.37 & (0.896, 19.7) & $-1.73$ & 1.80 & 309.8 & 0.999 & 0\\
%\hline
  & & Single & 0.070 & 16.4 & & & & 612.4 & 0.997 & 127\\
B5 $^{22}\mathrm{Na}$  & 0.05 & Double & 0.069 & 10.7 & 0.229 & & & 611.8 & 0.998 & 0\\
  & &  Multirate & 0.051 & 5.72 & (0.318, 8.21) & $-1.69$ & 2.10 & 610.1 & 0.998 & 13\\
%\hline
  & & Single & 0.071 & 14.3 & & & & 0.479 & 0.989 & 717\\
B7 $^{22}\mathrm{Na}$  & 0.5 & Double & 0.066 & 9.89 & 0.473 & & & 0.601 & 0.999 & 183\\
  & &  Multirate & 0.061 & 8.46 & (0.398, 0.314) & $-0.921$ & 0.831 & 0.580 & 0.999 & 0\\
%\hline
  & & Single & 0.068 & 12.9 & & & & 65.2 & 0.997 & 496\\
B8 $^{22}\mathrm{Na}$  & 0.5 & Double & 0.065 & 7.01 & 0.289 & & & 64.6 & 0.999 & 379\\
  & &  Multirate & 0.047 & 3.01 & (1.22, 45.8) & $-1.53$ & 1.86 & 64.8 & 1.000 & 0\\
%\hline
  & & Single & 0.068 & 15.8 & & & & see \textit{a} & 0.996 & 676\\
B\{4,5,8\} &  & Double & 0.066 & 9.38 & 0.230 & & & see \textit{b} & 0.998 & 75\\
  & &  Multirate & 0.045 & 3.46 & (1.48, 107.9) & $-1.57$ & 1.98 & see \textit{c} & 0.998 & 0\\
\hline
\end{tabular}
\textit{a}: $t_\mathrm{inj}$=\{305.8, 619.3, 65.0\} hours for B\{4,5,8\}\\
\textit{b}: $t_\mathrm{inj}$=\{306.6, 619.0, 65.3\} hours for B\{4,5,8\}\\
\textit{c}: $t_\mathrm{inj}$=\{305.2, 615.2, 64.9\} hours for B\{4,5,8\}
\end{table}

% Table 2
\begin{table}[ht]
\caption{PEST-Estimated Parameters Using $^{232}$U Breakthrough Data}
\label{tab:232U}
\begin{tabular}{|l|l|c|c|c|c|c|c|c|}
\hline
 Test & Model & $R_\mathrm{m}$ & $R_\mathrm{im}$ & $\omega_{D}$ & $\mu$ & $\sigma$& $t_\mathrm{inj}$ (hours) & $R^2$\\
\hline
   & Single & 3.65 & & & & & 1.92 & 0.946 \\
B3 & Double & 2.36 & 1.80 & 0.754 & & & 2.33 & 0.995 \\
   & Multirate & 1.63 & 5.68 & (1.44, 176.4) & $-1.86$ & 2.11 & 3.00 & 0.998 \\
%\hline
   & Single & 3.49 & & & & & 2.33 & 0.987 \\
B7 & Double & 3.52 & 2.83 & 0.022 & & & 2.15 & 0.993 \\
   & Multirate & 3.48 & 1.30 & (66.6, $2.45\times 10^7$) & $-2.21$ & 3.58 & 2.12 & 0.991 \\
\hline
\end{tabular}
\end{table}

\end{document}